\documentclass[prl,twocolumn,superscriptaddress]{revtex4-1}
\usepackage{}
\usepackage{amssymb}
\usepackage{amsfonts}
\usepackage{bbm}
\usepackage{mathrsfs}
\usepackage{graphics,graphicx,epsfig,bm,amsmath,amsthm,amssymb}
\usepackage{bm}
\usepackage{bbm}
\usepackage{longtable}
\usepackage{multirow}
\usepackage{array}
\usepackage{color}
\usepackage{diagbox}
\usepackage[usenames,dvipsnames]{xcolor}

\usepackage{float}
\usepackage[a4paper,colorlinks=true,
linkcolor=blue,citecolor=blue,
pdfauthor={ },
pdftitle={ },
pdfsubject={ },
pdfkeywords={ }]{hyperref}

\begin{document}

\title{Dynamically encircling an exceptional point in a real quantum system}

\author{Wenquan Liu}
\thanks{These authors contributed equally to this work.}
\affiliation{Hefei National Laboratory for Physical Sciences at the Microscale, University of Science and Technology of China, Hefei 230026, China}
\affiliation{CAS Key Laboratory of Microscale Magnetic Resonance and Department of Modern Physics, University of Science and Technology of China, Hefei 230026, China}
\affiliation{Synergetic Innovation Center of Quantum Information and Quantum Physics, University of Science and Technology of China, Hefei 230026, China}

\author{Yang Wu}
\thanks{These authors contributed equally to this work.}
\affiliation{Hefei National Laboratory for Physical Sciences at the Microscale, University of Science and Technology of China, Hefei 230026, China}
\affiliation{CAS Key Laboratory of Microscale Magnetic Resonance and Department of Modern Physics, University of Science and Technology of China, Hefei 230026, China}
\affiliation{Synergetic Innovation Center of Quantum Information and Quantum Physics, University of Science and Technology of China, Hefei 230026, China}

\author{Chang-Kui Duan}
\affiliation{Hefei National Laboratory for Physical Sciences at the Microscale, University of Science and Technology of China, Hefei 230026, China}
\affiliation{CAS Key Laboratory of Microscale Magnetic Resonance and Department of Modern Physics, University of Science and Technology of China, Hefei 230026, China}
\affiliation{Synergetic Innovation Center of Quantum Information and Quantum Physics, University of Science and Technology of China, Hefei 230026, China}

\author{Xing Rong}
\email{xrong@ustc.edu.cn}
\affiliation{Hefei National Laboratory for Physical Sciences at the Microscale, University of Science and Technology of China, Hefei 230026, China}
\affiliation{CAS Key Laboratory of Microscale Magnetic Resonance and Department of Modern Physics, University of Science and Technology of China, Hefei 230026, China}
\affiliation{Synergetic Innovation Center of Quantum Information and Quantum Physics, University of Science and Technology of China, Hefei 230026, China}

\author{Jiangfeng Du}
\email{djf@ustc.edu.cn}
\affiliation{Hefei National Laboratory for Physical Sciences at the Microscale, University of Science and Technology of China, Hefei 230026, China}
\affiliation{CAS Key Laboratory of Microscale Magnetic Resonance and Department of Modern Physics, University of Science and Technology of China, Hefei 230026, China}
\affiliation{Synergetic Innovation Center of Quantum Information and Quantum Physics, University of Science and Technology of China, Hefei 230026, China}

\begin{abstract}
%Non-Hermiticity dramatically alters the dynamics of physical systems and is refreshing our cognition of quantum physics\cite{NP_Ganainy, NPhoton_Feng}.
The exceptional point, known as the non-Hermitian degeneracy, has special topological structure, leading to various counterintuitive phenomena and novel applications\cite{NP_Ganainy, NPhoton_Feng,Science_Miri, Nat.Mater_Ozdemir}, which are refreshing our cognition of quantum physics.
One particularly intriguing behavior is the mode switch phenomenon induced by dynamically encircling an exceptional point in the parameter space\cite{PRL_Lefebvre, PRL_Atabek,JPA_Uzdin,JPA_Berry,PRA_Gilary,PRA_Thomas,PRL_Hassan,PRR_Pick}.
While these mode switches have been explored in classical systems\cite{Nature_Doppler, Nature_Xu, Nature_Yoon, PRX_Zhang}, the experimental investigation in the quantum regime remains elusive due to the difficulty of constructing time-dependent non-Hermitian Hamiltonians in a real quantum system.
Here we experimentally demonstrate dynamically encircling the exceptional point with a single nitrogen-vacancy center in diamond.
The time-dependent non-Hermitian Hamiltonians are realized utilizing a dilation method.
Both the asymmetric and symmetric mode switches have been observed.
Our work reveals the topological structure of the exceptional point and paves the way to comprehensively explore the exotic properties of non-Hermitian Hamiltonians in the quantum regime.
\end{abstract}

\maketitle

Quantum systems driven by non-Hermitian Hamiltonians exhibit exotic properties comparing to those governed by Hermitian Hamiltonians\cite{NP_Ganainy, NPhoton_Feng}.
One of the distinct features is the existence of exceptional point (EP), at which both the eigenvalues and the corresponding eigenvectors coalesce\cite{Czech.J.Phys_Berry}.
Classical systems, such as optics and photonics, have provided fertile grounds to investigate EP-related novel phenomena and applications\cite{Science_Miri, Nat.Mater_Ozdemir},
such as single mode lasers with gain and loss\cite{Science_feng,Science_Hodaei, Science_Peng}, unidirectional invisibility\cite{PRL_Lin,Nat.Mater_Feng}, EP-enhanced mode splitting\cite{PRL_Wiersig, Nature_Chen, Nature_Hodael,PRL_Liu} and many others\cite{Nature_Regensburger, PRL_Regensburger, NatMater_Weimann, NP_Peng}.
These fascinating processes were realized at or near an EP, but some other important properties of the EP can only be revealed when it is encircled in the parameter space.
When the EP is encircled in a quasistatic manner, the two eigenvalues and the corresponding eigenvectors swap with each other\cite{Eur.phys_Heiss, PRE_Heiss, PRL_Dembowski, Nature_Gao, PRX_Ding}.
The effects due to dynamically encircling EPs have been investigated in classical systems\cite{Nature_Doppler,Nature_Xu,Nature_Yoon,PRX_Zhang}.
Mode switching will emerge when the start points locate at different parity-time-symmetric ($\mathcal{PT}$-symmetric) phases\cite{PRL_Lefebvre, PRL_Atabek,JPA_Uzdin,JPA_Berry,PRA_Gilary,PRA_Thomas,PRL_Hassan,PRR_Pick}.
These mode switches are expected to play an important role in quantum control\cite{PRR_Pick}.

\begin{figure*}[http]
\centering
\includegraphics[width=2.0\columnwidth]{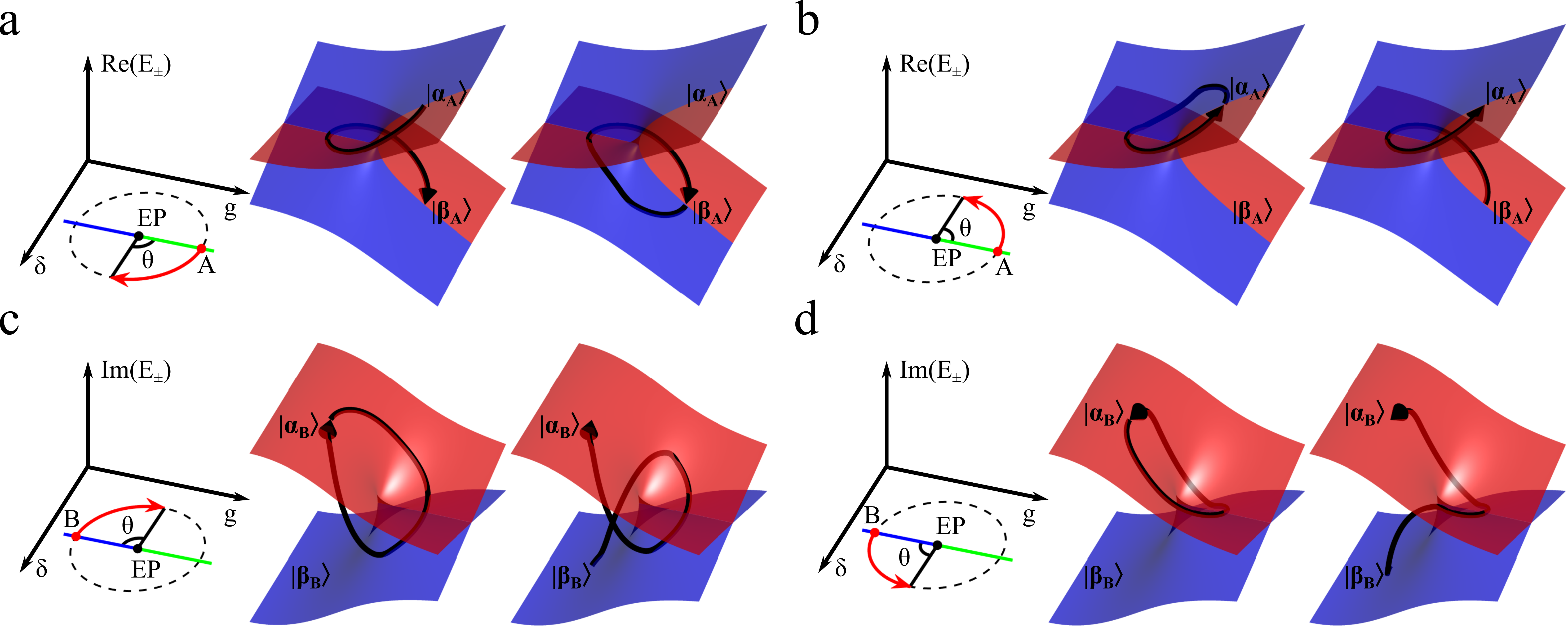}
\caption{\textbf{Asymmetric and symmetric mode switching by dynamically encircling the EP.}
\textbf{a-d}, Encircling paths in the parameter space and prospective encircling trajectories in the eigenvalue Riemann sheets when the start point locates at $\mathcal{PT}$-symmetric phase (a,b) and $\mathcal{PT}$-symmetry broken phase (c,d).
The left panel in each diagram shows the encircling path.
Green line denotes the Hamiltonian is in $\mathcal{PT}$-symmetric phase and blue line means the Hamiltonian is in $\mathcal{PT}$-symmetry broken phase.
Red arc with arrow shows the encircling direction.
The middle and right panel in each diagram displays the prospective encircling trajectory.
The encircling initial states in the middle panel are $|\alpha_A\rangle$ (a,b) and $|\alpha_B\rangle$ (c,d) while the encircling initial states in the right panel are $|\beta_A\rangle$ (a,b) and $|\beta_B\rangle$ (c,d).
$|\alpha_A\rangle$ and $|\beta_A\rangle$ ($|\alpha_B\rangle$ and $|\beta_B\rangle$) are eigenstates of the initial Hamiltonian at start point A (B).
}
\label{Fig1}
\end{figure*}

We take the following non-Hermitian Hamiltonian as an example to describe the mode switching phenomenon:
\begin{equation}
\label{System Hamiltonian}
H_s = \left( \begin{array}{cc}
\delta/2+i\gamma & g\\
g  & -\delta/2-i\gamma
\end{array}
\right ),
\end{equation}
where $\gamma$ is a constant number chosen as $\gamma=1$,  $\delta$ and $g$ are time-dependent real numbers.
Eigenenergies of this Hamiltonian are $E_\pm=\pm\sqrt{g^2+\delta^2/4-1+i\delta}$, and a pair of EPs arises when $\delta=0$ and $g=\pm1$.
For different $\delta$ and $g$, the real and imaginary parts of $E_\pm$ are shown in Fig.~\ref{Fig1}, which displays a complex eigenvalue topology of two intersecting Riemann sheets wrapped around the EP.
Dynamically encircling the EP in the parameter space is realized by the time-dependent parameters, $\delta(t)=0.5\sin[\theta(t)+\theta_0]$ and $g(t)=1+0.5\cos[\theta(t)+\theta_0]$, where $\theta(t)=\omega t$ is the encircling angle and $\theta_0$ defines the start point.
The encircling direction is decided by the sign of $\omega$ as shown by the red arc with arrow.
In Fig.~\ref{Fig1}a and b, $\theta_0=0$ and the start points locate at the $\mathcal{PT}$-symmetric phase.
The encircling initial state is prepared to $|\alpha_A\rangle$ or $|\beta_A\rangle$ which is the eigenstate of the initial Hamiltonian at start point A.
When the encircling direction is clockwise, both $|\alpha_A\rangle$ and $|\beta_A\rangle$ will evolve to $|\beta_A\rangle$ (Fig.~\ref{Fig1}a).
When the encircling direction is counterclockwise, both $|\alpha_A\rangle$ and $|\beta_A\rangle$ will evolve to $|\alpha_A\rangle$ (Fig.~\ref{Fig1}b).
The encircling final state depends on the encircling direction, which is taken as the asymmetric mode switching.
In Fig.~\ref{Fig1}c and d, $\theta_0=\pi$ and the start points locate at the $\mathcal{PT}$-symmetry broken phase.
The encircling initial state is prepared to $|\alpha_B\rangle$ or $|\beta_B\rangle$ which is eigenstate of the initial Hamiltonian at start point B.
Whether the encircling direction is clockwise (Fig.~\ref{Fig1}c) or counterclockwise (Fig.~\ref{Fig1}d), both $|\alpha_B\rangle$ and $|\beta_B\rangle$ will evolve to $|\alpha_B\rangle$. This exhibits the symmetric mode switching.

Experimental investigations of phenomena via dynamically encircling EPs have been realized in classical systems\cite{Nature_Doppler,Nature_Xu,Nature_Yoon,PRX_Zhang}, and the research has remained elusive in the quantum domain.
This is because the Hamiltonian of a closed quantum system is Hermitian while the evolution of dynamically encircling an EP is governed by the time-dependent non-Hermitian Hamiltonian.
Recently, there have been experiments investigating dynamics of individual quantum systems driven by non-Hermitian Hamiltonians\cite{Science_Wu, NP_M.Naghiloo, PRB_Partanen}. In these works, the Hamiltonians are time-independent, which prevent further investigating important physics driven by time-dependent non-Hermitian Hamiltonians.
To overcome this obstacle, we have utilized  a dilation method to realize time-dependent  non-Hermitian Hamiltonians. Then dynamically encircling an EP has been realized in a single-spin system and mode switches have been observed.

The procedure of dynamically encircling an EP is described by the Schr$\ddot{\mathrm{o}}$dinger equation $i\partial_t|\psi(t)\rangle=H_s(t)|\psi(t)\rangle$.
We introduce an ancilla qubit to construct the dilated state  $|\Psi(t)\rangle=|\psi(t)\rangle\otimes|-\rangle+\eta(t)|\psi(t)\rangle\otimes|+\rangle$ (unnormalized for convenience), where $|-\rangle=(|0\rangle-i|1\rangle)/\sqrt2$ and $|+\rangle=-i(|0\rangle+i|1\rangle)/\sqrt2$ form an orthogonal basis of the ancilla qubit, and $\eta(t)$ is an appropriate operator given by the dilation method (see Supplementary Note \uppercase\expandafter{\romannumeral2} for details).
The evolution of $|\psi(t)\rangle$ is embedded in the subspace of $|\Psi(t)\rangle$ where the state of the ancilla qubit is $|-\rangle$.
The state $|\Psi(t)\rangle$ is governed by the Hermitian Hamiltonian, $H_{s,a}(t)$, which can be flexibly designed according to practical quantum systems.
Here we choose
\begin{equation}
H_{s,a}(t) = \Lambda(t) \otimes I + \Gamma(t) \otimes \sigma_z,
\label{Dilated Hamiltonian_Main_Text}
\end{equation}
where $\Lambda(t)=\{H_s(t)+ib(t) I+[i\frac{d}{dt}\eta(t)+\eta(t)H_s(t)+ib(t)\eta(t)]\eta(t)\}M^{-1}(t)$, $\Gamma(t)=i[H_s(t)\eta(t)-\eta(t)H_s(t)-i\frac{d}{dt}\eta(t)]M^{-1}(t)$, $M(t)=\eta^{\dagger}(t)\eta(t)+I$ and $b(t)$ is a real function (see Supplementary Note \uppercase\expandafter{\romannumeral2} for details).
During the encircling process, the inner product of the encircling state, $\langle\psi(t)|\psi(t)\rangle$, covers several orders of magnitude since the evolution under  the non-Hermitian Hamiltonian $H_s(t)$ is trace non-conservative.
The probability of obtaining the ancilla qubit state $|-\rangle$, $P_{-}(t)$, will be small during some periods of evolution, which makes it difficult to continuously monitor the state evolution of dynamically encircling the EP though selecting the state $|-\rangle$ of the ancilla qubit  (see Supplementary Note \uppercase\expandafter{\romannumeral3} for details).
To solve this problem, we introduce an alternative measurement method.
The dilated state can also be rewritten as $|\Psi(t)\rangle=[I-i\eta(t)]|\psi(t)\rangle\otimes|0\rangle-i[I+i\eta(t)]|\psi(t)\rangle\otimes|1\rangle$.
The state of the system qubit is measured when the ancilla qubit state is $|0\rangle$, which is $|\chi(t)\rangle=[I-i\eta(t)]|\psi(t)\rangle$.
Then the state $|\psi(t)\rangle$ is obtained by multiplying $[I-i\eta(t)]^{-1}$ to the state $|\chi(t)\rangle$ with proper renormalization.

\begin{figure}[http]
\centering
\includegraphics[width=1.0\columnwidth]{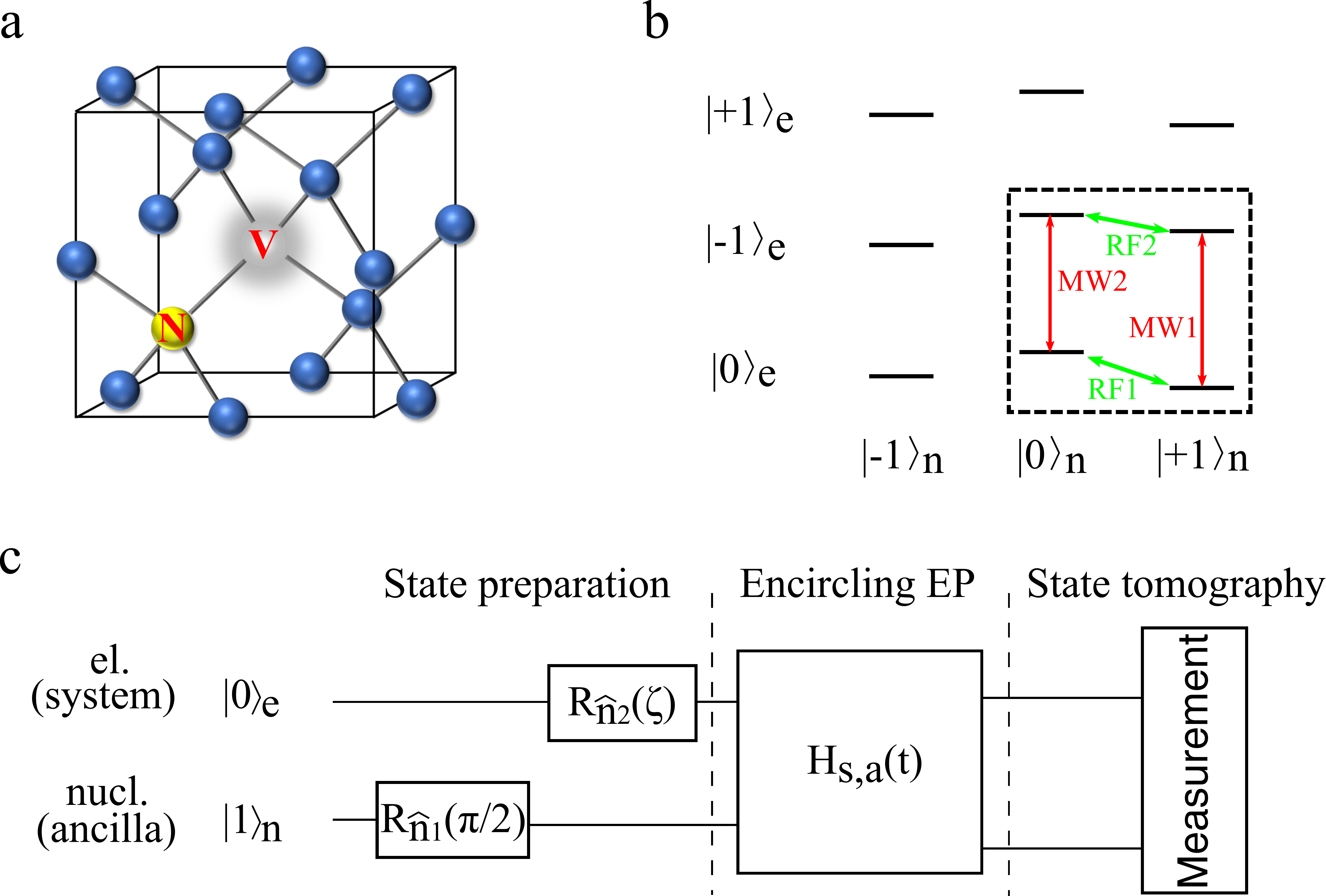}
\caption{\textbf{Realization of dynamically encircling the EP in a NV center.}
\textbf{a}, Atomic structure of the NV center.
\textbf{b}, The hyperfine energy levels of the coupling system with NV electron spin and $^{14}\mathrm{N}$ nuclear spin. The four energy levels in the black box is utilized to form a two-qubit system.
\textbf{c}, Quantum circuit of the experiment.
The coupling system is initialized to the state $|0\rangle_e|1\rangle_n$.
Rotation $\pi/2$ along axis $\widehat{n}_1$ on the nuclear spin followed by rotation $\zeta$ along axis $\widehat{n}_2$ on the electron spin can prepare the coupling system to different initial states.
Dynamically encircling the EP is realized by adding two selective microwave pulses (red arrows in (b)). Then state tomography is implemented to measure the state $|\chi(t)\rangle$ for further obtainment of the state $|\psi(t)\rangle$.
}
\label{Fig2}
\end{figure}

Our scheme is demonstrated in a single nitrogen-vacancy ($\mathrm{NV}$) center in diamond (Fig.~\ref{Fig2}a).
With a static magnetic field applied along the NV symmetry axis, the Hamiltonian of the $\mathrm{NV}$ center can be written as $H_{\mathrm{NV}} = 2\pi(DS_z^2 + \omega_eS_z + QI_z^2 + \omega_nI_z + AS_zI_z)$,
where $D=2.87$ GHz is the zero-field splitting of the electron spin, $Q=-4.95$ MHz is the nuclear quadrupolar interaction, and $A=-2.16$ MHz is the hyperfine coupling between the electron spin and the nuclear spin.
$\omega_e$ ($\omega_n$) denotes the Zeeman splitting of the electron (nuclear) spin.
$S_z$ and $I_z$ are the spin-1 operators of the electron spin and the nuclear spin, respectively.
The subspace spanned by $|0\rangle_e|1\rangle_n, |0\rangle_e|0\rangle_n, |-1\rangle_e|1\rangle_n$ and $|-1\rangle_e|0\rangle_n$ (black box in Fig.~\ref{Fig2}b) is encoded as a two-qubit system to construct Hamiltonian $H_{s,a}(t)$.
The electron spin is chosen as the system qubit and the nuclear spin is selected as the ancilla qubit.
By decomposing $\Lambda(t)$ and $\Gamma(t)$ in terms of Pauli operators, we can rewrite the Hamiltonian $H_{s,a}(t)$ in Eq.\ref{Dilated Hamiltonian_Main_Text} as $H_{s,a}(t)=\sum_{i=0}^3[ A_i(t) \sigma_i\otimes I+B_i(t)\sigma_i\otimes \sigma_z]$, where $A_i(t)=\mathrm{Tr}[\Lambda(t)\cdot \sigma_i]/2$, $B_i(t)=\mathrm{Tr}[\Gamma(t)\cdot \sigma_i]/2$ are the decomposition coefficients,  and $\sigma_i=I, \sigma_x,\sigma_y,\sigma_z$ for $i=0,1,2,3$ are Pauli operators.
Two selective microwave pulses (red arrows in Fig.~\ref{Fig2}b) are applied to realize the Hamiltonian $H_{s,a}(t)$.
The Hamiltonian of the pulses in the two-qubit system can be written as
\begin{equation}
\label{MW Pulses}
\begin{aligned}
H_c
= &\Omega_1(t)\cos[\int_0^t\omega_1(\tau)d\tau+\phi_1(t)]\sigma_x\otimes|1\rangle_n~_n\langle1|\\
&+\Omega_2(t)\cos[\int_0^t\omega_2(\tau)d\tau+\phi_2(t)]\sigma_x\otimes|0\rangle_n~_n\langle0|.
\end{aligned}
\end{equation}
The dilated Hamiltonian $H_{s,a}(t)$ can be constructed in an interaction picture when the amplitudes, angular frequencies and the phases of the two pulses satisfy the following relations,
$\Omega_{1,2}(t)=2\sqrt{[A_1(t)\pm B_1(t)]^2+[A_2(t)\pm B_2(t)]^2}$,
$\omega_{1,2}(t)=\omega_{\mathrm{MW}1,2}+2A_3(t)\pm2B_3(t)$,
and $\phi_{1,2}(t)=-\mathrm{arctan2}[A_2(t)\pm B_2(t),A_1(t)\pm B_1(t)]$
(see Supplementary Note \uppercase\expandafter{\romannumeral4} for details).

The experiment was implemented on an optically detected magnetic resonance setup.
The static magnetic field was set to 500 Gauss in order to polarize the NV center to state $|0\rangle_e|1\rangle_n$ by optical pumping\cite{PRL_V.Jacques}.
When we choose $\eta(0)=\eta_0\cdot I$, the initial state of the two-qubit system has the form $|\Psi(0)\rangle=|\psi(0)\rangle\otimes(|-\rangle+\eta_0|+\rangle)$.
The encircling initial state $|\psi(0)\rangle$ was set to the eigenstate, $|\alpha_{A}\rangle$ or $|\beta_{A}\rangle$ ($|\alpha_{B}\rangle$ or $|\beta_{B}\rangle$), of the initial Hamiltonian at start point A (B).
$|\Psi(0)\rangle$ was prepared by rotation $R_{\widehat{n}_1}(\pi/2)$ on the nuclear spin followed by rotation $R_{\widehat{n}_2}(\zeta)$ on the electron spin (Fig.~\ref{Fig2}c), where $\widehat{n}_1=(1-\eta_0^2,2\eta_0,0)/(1+\eta_0^2)$, $\widehat{n}_2$ and $\zeta$ depended on the choice of $|\psi(0)\rangle$.
Then the system evolved under the dilated Hamiltonian $H_{s,a}(t)$ by applying two selective $\mathrm{MW}$ pulses given in Eq.\ref{MW Pulses} to realize the dynamically encircling procedure.
The total encircling time is set to be $T=15~\mu s$.
Finally, a state tomography is implemented to measure the state $|\chi(t)\rangle$ when the ancilla qubit state is $|0\rangle$.
The state $|\psi(t)\rangle$ is obtained by multiplying the operator $[I-i\eta(t)]^{-1}$ on the state $|\chi(t)\rangle$ with the maximum likelihood estimation\cite{PRA_Fiurasek} and all the experimental fidelities of $|\chi(t)\rangle$ and $|\psi(t)\rangle$ in our experiment are close to 1 (see Supplementary Note \uppercase\expandafter{\romannumeral3} for details).

\begin{figure}[http]
\centering
\includegraphics[width=1\columnwidth]{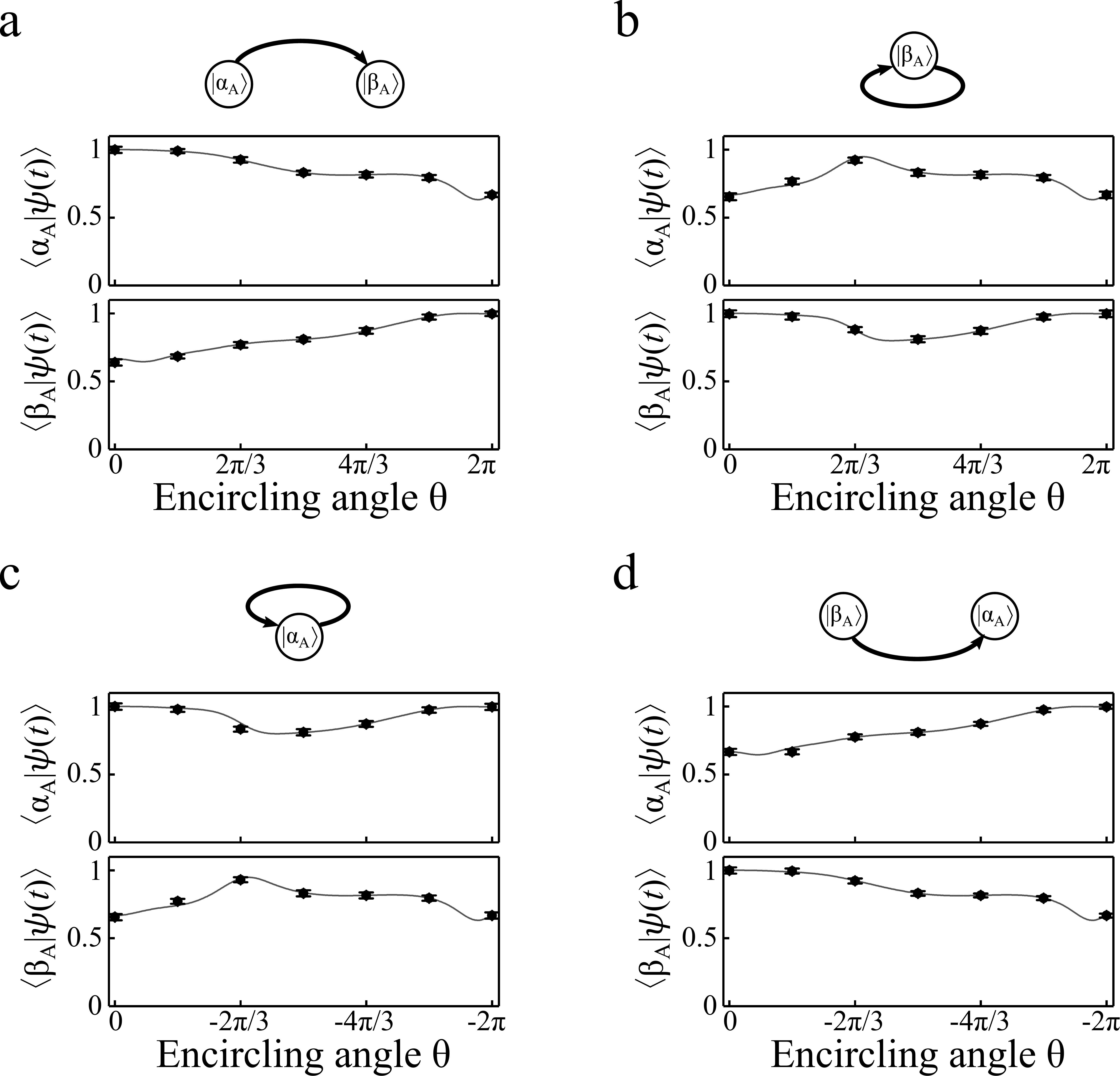}
\caption{\textbf{Asymmetric mode switching by starting from $\mathcal{PT}$-symmetric phase.}
\textbf{a-b}, Clockwise encircling the EP starting from eigenstates $|\alpha_A\rangle$ (a) and $|\beta_A\rangle$ (b).
\textbf{c-d}, Counterclockwise encircling the EP starting from eigenstates $|\alpha_A\rangle$ (c) and $|\beta_A\rangle$ (d).
The evolution of the encircling state is characterized by the overlaps $\langle\alpha_A|\psi(t)\rangle$ (upper panel in each diagram) and $\langle\beta_A|\psi(t)\rangle$ (bottom panel in each diagram).
Black dots  with error bars are experimental results, and grey lines are the simulation predications.
}
\label{Fig3}
\end{figure}

\begin{figure}[http]
\centering
\includegraphics[width=1\columnwidth]{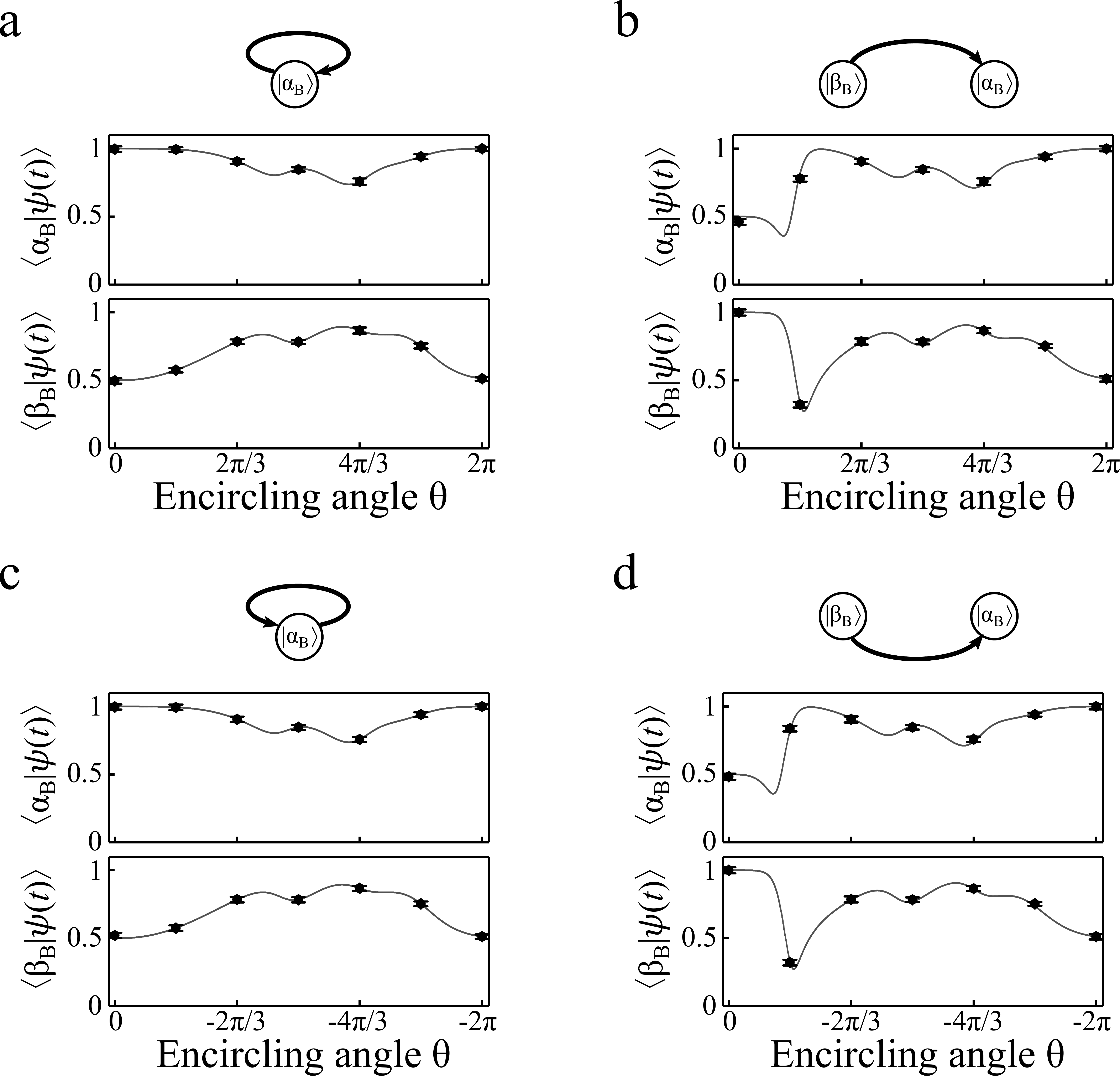}
\caption{\textbf{Symmetric mode switching by starting from $\mathcal{PT}$-symmetry broken phase.}
\textbf{a-b}, Clockwise encircling the EP starting from eigenstates $|\alpha_B\rangle$ (a) and $|\beta_B\rangle$ (b).
\textbf{c-d}, Counterclockwise encircling the EP starting from eigenstates $|\alpha_B\rangle$ (c) and $|\beta_B\rangle$ (d).
The evolution of the encircling state is characterized by the overlaps $\langle\alpha_B|\psi(t)\rangle$ (upper panel in each diagram) and $\langle\beta_B|\psi(t)\rangle$ (bottom panel in each diagram).
Black dots  with error bars are experimental results, and grey lines are the simulation predications.
}
\label{Fig4}
\end{figure}

Fig.~\ref{Fig3} displays the results of an asymmetric mode switching when the start point locates at the $\mathcal{PT}$-symmetric phase.
The horizontal axes in Fig.~\ref{Fig3} and \ref{Fig4} are rotation angle $\theta$ for clearly demonstration of the encircling process, where $\theta = \omega t$ with $\omega =\pm 2\pi/15~ \mathrm{rad}\cdot\mu$s$^{-1}$.
Fig.~\ref{Fig3}a and b show the cases of clockwise encircling.
In Fig.~\ref{Fig3}a, we prepared the encircling initial state, $|\psi(0)\rangle$, to eigenstate $|\alpha_A\rangle$.
The overlaps of $|\psi(0)\rangle$ with $|\alpha_A\rangle$ and $|\beta_A\rangle$ were $1.00(2)$ and $0.64(2)$, respectively.
The non-zero overlap between $|\psi(0)\rangle$ and $|\beta_A\rangle$ is due to the non-orthogonality of the two eigenstates of $H_s(0)$.
Then the encircling state followed the evolution of dynamically encircling the EP characterized by the overlaps with $|\alpha_A\rangle$ and $|\beta_A\rangle$.
After the encircling process, the overlap between the encircling final state and $|\beta_A\rangle$ was $1.00(2)$.
This shows that eigenstate $|\alpha_A\rangle$ will switch to $|\beta_A\rangle$ after encircling around the EP, as clarified by the sketch on the diagram.
In Fig.~\ref{Fig3}b, the state was first prepared to $|\beta_A\rangle$ and the results show that it returned to $|\beta_A\rangle$ after the encircling process.
The counterclockwise encircling cases are shown in Fig.~\ref{Fig3}c and d.
The initial states were prepared to $|\alpha_A\rangle$ and $|\beta_A\rangle$ in Fig.~\ref{Fig3}c and d, respectively.
After encircling the EP, the encircling states both evolved to $|\alpha_A\rangle$.
All the measured evolutions in Fig.~\ref{Fig3} show good agreement with the theoretical predictions.
These results exhibit that the encircling initial state doesn't influence the encircling final state.
The encircling final state depends on the encircling direction.
The evolutions displayed in Fig.~\ref{Fig3} unambiguously certifies an asymmetric mode switching when the evolution starts from the $\mathcal{PT}$-symmetric phase.

Fig.~\ref{Fig4} exhibits the results of a symmetric mode switching when the start point locates at the $\mathcal{PT}$-symmetry broken phase.
The overlaps between the encircling state and the eigenstates, $|\alpha_B\rangle$ and $|\beta_B\rangle$, were used to describe the evolution.
Fig.~\ref{Fig4}a and b show the cases of clockwise encircling and the initial states were prepared to $|\alpha_B\rangle$ and $|\beta_B\rangle$, respectively.
The counterclockwise encircling cases are presented in Fig.~\ref{Fig4}c and d.
For all these cases, the final states evolved to $|\alpha_B\rangle$ after the encircling process regardless of the initial state and the path direction.
The experimental results agree well with the corresponding theoretical predictions.
These results indisputably reveal a symmetric mode switching when initial Hamiltonian is $\mathcal{PT}$-symmetry broken.

In conclusion, dynamically encircling an EP in a single-spin system has been realized to explore the topological structure of the EP, and both the symmetric and asymmetric mode conversions have been observed. These mode switches provide a robust fashion to control quantum states and thus have great potential in quantum information precessing\cite{PRR_Pick}. Furthermore, our successful demonstration of  engineering time-dependent non-Hermitian Hamiltonians opens a door towards  future investigation of complicated dynamical processes governed by time-dependent non-Hermitian Hamiltonians in the quantum realm,
such as, encircling high-order EPs\cite{J.Phys.A_Demange, PRA_Schnabel}, researching non-Hermitian geometric phase\cite{J.Math.Phys_Hossein, PRA_Gong} and studying non-Hermitian topological invariants\cite{J.Phys_Ananya}.

\subsection{Acknowledgements}
We thank Qing-Hai Wang for the helpful discussion. This work was supported by the National Key R$\&$D Program of China (Grants No. 2018YFA0306600 and No. 2016YFB0501603), the NNSFC (No. 11761131011), the Chinese Academy of Sciences (Grants No. GJJSTD20170001, No.QYZDY-SSW-SLH004 and No.QYZDB-SSW-SLH005), and Anhui Initiative in Quantum Information Technologies (Grant No. AHY050000). X.R. thanks the Youth Innovation Promotion Association of Chinese Academy of Sciences for the support.
%
% J.D. and X. R. proposed the idea and supervised the experiments. W.L., Y. W. and X. R. designed and performed experiments.  W.L. and C.-K.D. wrote the paper. All authors analyzed the data, discussed the results and commented on the manuscript.
%
%The authors declare no competing interests.
%
%All data are available in the manuscript or the supplementary materials.
%
%Correspondence and requests for materials should be addressed to J.D. (djf@ustc.edu.cn) and X. R. (xrong@ustc.edu.cn).

\subsection{Method}

Our experiment was implemented on a NV center in a piece of $[100]$ face bulk diamond which is isotopically purified ([$^{12}$C]=99.9\%). The electron spin qubit was chosen as the system qubit. The dephasing time $T_2^\star$ of the electron spin is $36(3) \mathrm{\mu s}$  (see Supplementary Note \uppercase\expandafter{\romannumeral1} for details).

The sample was placed in a confocal setup.
An acousto-optic modulator (ISOMET, AOMO 3200-121) was utilized to modulate the 532 nm green laser pulses.
The laser beam traveled twice through the acousto-optic modulator before going through an oil objective (Olympus, PLAPON 60*O, NA 1.42) and then focusing on the NV center.
The phonon sideband fluorescence (wavelength, 650-800nm) went through the same oil objective and was collected by an avalanche photodiode (Perkin Elmer, SPCM-AQRH-14) with a counter card.
The static magnetic field of 500 G was provided by a permanent magnet along the NV symmetry axis.
 The state of the two-qubit system can be effectively polarized to state $|0\rangle_e|1\rangle_n$ by laser pumping.
Microwave (MW) and radio-frequency (RF) pulses generated by an arbitrary waveform generator (Keysight, M8190A)  were applied to manipulate the state of the two-qubit system.
The MW pulses were amplified by an amplifier (AmpliTech, APTMP3-01001800-2520-D4) and fed by a coplanar waveguide.
The RF pulses were carried by a RF coil after the amplification of a power amplifier (Mini-Circuits, LZY-22+).

\newpage

\onecolumngrid
\vspace{1.5cm}
\begin{center}
\textbf{\large Supplementary Material}
\end{center}

\setcounter{figure}{0}
\setcounter{equation}{0}
\setcounter{table}{0}
\makeatletter
\renewcommand{\thefigure}{S\arabic{figure}}
\renewcommand{\theequation}{S\arabic{equation}}
\renewcommand{\thetable}{S\arabic{table}}
\renewcommand{\bibnumfmt}[1]{[RefS#1]}
\renewcommand{\citenumfont}[1]{RefS#1}

\section{Note \uppercase\expandafter{\romannumeral1}: Sample}
\quad \ \ Our experiment was implemented on a NV center in $[100]$ face bulk diamond which was isotopically purified ([$^{12}$C]=99.9\%). The electron spin qubit was chosen as the system qubit. The dephasing time, $T_2^\star=36(3) \mathrm{\mu s}$, of the electron spin is obtained by Ramsey sequence as shown in Supplementary Fig.~\ref{Dephasing time of the NV center}.

\begin{figure}[htbp]
\centering
\includegraphics[width=15cm]{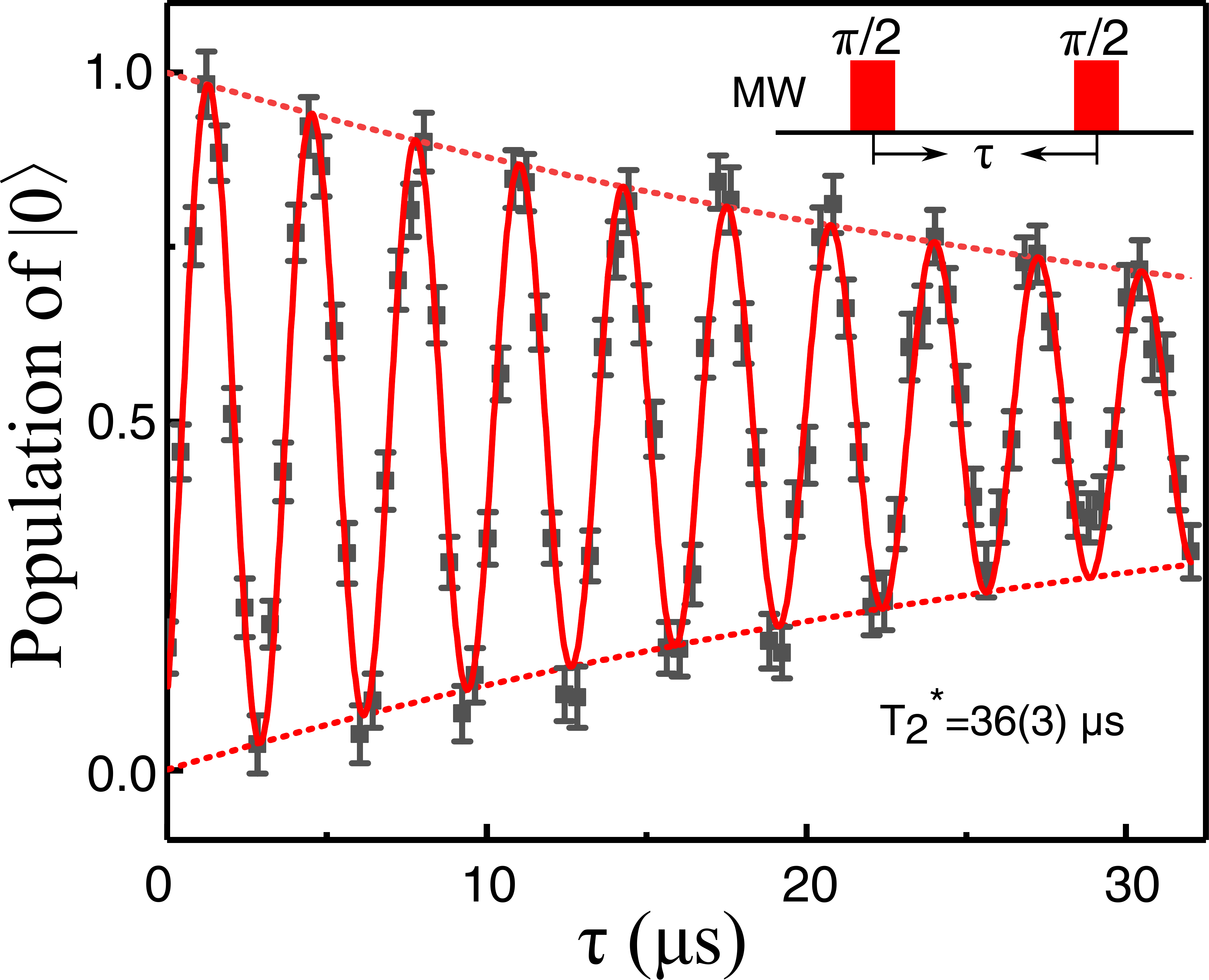}
\caption{\textbf{Dephasing time of the electron spin.}   Result of the Ramsey experiment (insert, pulse sequence) for the electron spin. The solid red line is the fit to the experiment data (black square), the red dashed line is the fit to the envelope curve. The error bars on the data points are the standard deviations from the mean.}
\label{Dephasing time of the NV center}
\end{figure}

\section{Note \uppercase\expandafter{\romannumeral2}: Dilation method}
\quad ~This part gives the derivation of how the dilated Hamiltonian $H_{s,a}(t)$ in equation (2) of the main text is obtained.
%#####%  Dilation Method---Description
The evolution of dynamically encircling the EP is described by the Schr$\ddot{\mathrm{o}}$dinger equation:
\begin{equation}
\label{DEEP Schrodinger Equation}
i\partial_t|\psi(t)\rangle=H_s(t)|\psi(t)\rangle.
\end{equation}
The non-unitary evolution of the state $|\psi(t)\rangle$ is realized in a quantum system by a dilation method.
We introduce an ancilla qubit and construct the following dilated state
\begin{equation}
\label{Dilated State}
|\Psi(t)\rangle=\frac{|\psi(t)\rangle\otimes|-\rangle+\eta(t)|\psi(t)\rangle\otimes|
+\rangle}{\sqrt{\langle\psi(t)|\eta^\dagger(t)\eta(t)+I|\psi(t)\rangle}},
\end{equation}
where $|-\rangle=(|0\rangle-i|1\rangle)/\sqrt2$, $|+\rangle=-i(|0\rangle+i|1\rangle)/\sqrt2$ are the eigenstates of Pauli operator $\sigma_y$, which form an orthogonal basis of the ancilla qubit, and $\eta(t)$ is an appropriate operator.
The evolution of $|\psi(t)\rangle$ is embedded in the subspace of $|\Psi(t)\rangle$ where the ancilla qubit state is $|-\rangle$.
The state $|\Psi(t)\rangle$ is governed by the dilated Hermitian Hamiltonian $H_{s,a}(t)$ and its evolution follows the Schr$\ddot{\mathrm{o}}$dinger equation
\begin{equation}
\label{dilated Schrodinger Equation}
i\partial_t|\Psi(t)\rangle=H_{s,a}(t)|\Psi(t)\rangle.
\end{equation}
Once the form of the dilated Hamiltonian $H_{s,a}(t)$ is obtained, we can realize dynamically encircling the EP in a quantum system.

%#####%  Dilation Method---Solution
From equation \ref{DEEP Schrodinger Equation}, the solution of the encircling state is
\begin{equation}
\label{DEEP State Solution}
|\psi(t)\rangle=\mathcal{T}e^{-i\int_0^t H_s(\tau)d\tau}|\psi(0)\rangle      ~~~~~
\langle\psi(t)|=\langle\psi(0)|\widetilde{\mathcal{T}}e^{i\int_0^t H_s^\dagger(\tau) d\tau}.
\end{equation}
We define an operator $M(t)$ as
\begin{equation}
\label{operator M}
M(t)\equiv\mathcal{T}e^{-i\int_0^t [H_s(\tau)+ib(\tau)\cdot I]^\dagger d\tau}M(0)\widetilde{\mathcal{T}}e^{i\int_0^t [H_s(\tau)+ib(\tau)\cdot I] d\tau},
\end{equation}
where $b(t)$ is a real function and $M(0)$ is the initial operator of $M(t)$.
The relation between operator $M(t)$ and $\eta(t)$ is defined as
\begin{equation}
\label{Relation of operator  M and eta}
M(t)=\eta^{\dagger}(t)\eta(t)+I.
\end{equation}
With equation \ref{DEEP State Solution} and \ref{operator M}, the denominator of the dilated state in equation \ref{Dilated State} can be reduced to
\begin{equation}
\label{denominator simplification}
\begin{aligned}
&\sqrt{\langle\psi(t)|\eta^\dagger(t)\eta(t)+I|\psi(t)\rangle}                                                                                                    \\
=&\sqrt{\langle\psi(0)|\widetilde{\mathcal{T}}e^{i\int_0^t H_s^\dagger(\tau) d\tau}\mathcal{T}e^{-i\int_0^t [H_s(\tau)+ib(\tau)\cdot I]^\dagger d\tau}M(0)\widetilde{\mathcal{T}}e^{i\int_0^t [H_s(\tau)+ib(\tau)\cdot I] d\tau}\mathcal{T}e^{-i\int_0^t H_s(\tau)d\tau}|\psi(0)\rangle}                                                                                            \\
=&\sqrt{\langle\psi(0)|e^{-\int_0^tb(\tau)d\tau}M(0)e^{-\int_0^tb(\tau)d\tau}|\psi(0)\rangle}                                    \\
%=&e^{-\int_0^tb(\tau)d\tau}\sqrt{\langle\psi(0)|M(0)|\psi(0)\rangle}                                                                                 \\
=&e^{-\int_0^tb(\tau)d\tau}
\end{aligned},
\end{equation}
where we have chosen $M(0)=I/\langle\psi(0)|\psi(0)\rangle$.
Then the dilated state takes the form
\begin{equation}
\label{dilated state simplification}
\begin{aligned}
|\Psi(t)\rangle
&=e^{\int_0^tb(\tau)d(\tau)}[|\psi(t)\rangle\otimes|-\rangle+\eta(t)|\psi(t)\rangle\otimes|+\rangle]          \\
&=e^{-i\int_0^tib(\tau)\cdot Id\tau}[\mathcal{T}e^{-i\int_0^t H_s(\tau)d\tau}|\psi(0)\rangle\otimes|-\rangle+\eta(t)\mathcal{T}e^{-i\int_0^t H_s(\tau)d\tau}|\psi(0)\rangle\otimes|+\rangle]                                                   \\
%&=\mathcal{T}e^{-i\int_0^t [H_s(\tau)+ib(\tau)\cdot I]d\tau}|\psi(0)\rangle\otimes|-\rangle+\eta(t)\mathcal{T}e^{-i\int_0^t [H_s(\tau)+ib(\tau)\cdot I]d\tau}|\psi(0)\rangle\otimes|+\rangle                                                     \\
&=\mathcal{T}e^{-i\int_0^t H_s^\prime(\tau)d\tau}|\psi(0)\rangle\otimes|-\rangle+\eta(t)\mathcal{T}e^{-i\int_0^t H_s^\prime(\tau)d\tau}|\psi(0)\rangle\otimes|+\rangle,
\end{aligned}
\end{equation}
where we have defined $H_s^\prime(t)=H_s(t)+ib(t)\cdot I$, and equation \ref{DEEP State Solution} has been taken into account.
If we choose $|\psi^\prime(0)\rangle=|\psi(0)\rangle$, then $|\psi^\prime(t)\rangle=\mathcal{T}e^{-i\int_0^t H_s^\prime(\tau)d\tau}|\psi(0)\rangle$ is the solution of the Schr$\ddot{\mathrm{o}}$dinger equation
\begin{equation}
\label{schrodinger equation adding trick term}
i\partial_t|\psi^\prime(t)\rangle=H_s^\prime(t)|\psi^\prime(t)\rangle.
\end{equation}
In this condition, equation \ref{dilated state simplification} reduces to
\begin{equation}
\label{Simplified Dilated State}
|\Psi(t)\rangle=|\psi^\prime(t)\rangle\otimes|-\rangle+\eta(t)|\psi^\prime(t)\rangle\otimes|+\rangle.
\end{equation}
Comparing equations \ref{dilated Schrodinger Equation}, \ref{schrodinger equation adding trick term} and \ref{Simplified Dilated State} with equations 1-3 in the supplementary material of \cite{Science_Wu_SM},
we find that $H_{s,a}(t)$ can be obtained by dilating $H_s^\prime(t)$ utilize the dilation method in \cite{Science_Wu_SM}, as the definition of operator $\eta(t)$ and $M(t)$ here are the same as the ones in \cite{Science_Wu_SM}.
Considering the solution given in equation 20-21 in the supplementary material of \cite{Science_Wu_SM}, we obtain the solution of $H_{s,a}(t)$ have the form
 \begin{equation}
H_{s,a}(t) = \Lambda(t) \otimes I + \Gamma(t) \otimes \sigma_z,
\label{Dilated Hamiltonian}
\end{equation}
where
\begin{equation}
\label{Operator Lambda and Gamma}
\left\{
\begin{aligned}
\Lambda(t)&=\{H_s^\prime(t)+[i\frac{d}{dt}\eta(t)+\eta(t)H_s^\prime(t)]\eta(t)\}M^{-1}(t)   \\
&=\{H_s(t)+ib(t)\cdot I+[i\frac{d}{dt}\eta(t)+\eta(t)H_s(t)+ib(t)\eta(t)]\eta(t)\}M^{-1}(t),  \\
\Gamma(t)&=i[H_s^\prime(t)\eta(t)-\eta(t)H_s^\prime(t)-i\frac{d}{dt}\eta(t)]M^{-1}(t)           \\
&=i[H_s(t)\eta(t)-\eta(t)H_s(t)-i\frac{d}{dt}\eta(t)]M^{-1}(t),
\end{aligned}
\right.
\end{equation}
which is equation (2) in the main text.

\section{Note \uppercase\expandafter{\romannumeral3}: Measurement method}

\quad~This part explains the difficulty of obtaining the encircling state $|\psi(t)\rangle$ though measuring the system qubit state when the ancilla qubit state is $|-\rangle$ and shows how we measured $|\psi(t)\rangle$ experimentally.

%#####%  Dilation Method---Parameter Choose
Without loss of generality, two cases are chosen as examples for demonstration.
The encircling initial state is $|\alpha_B\rangle$ and $|\beta_B\rangle$ for case \uppercase\expandafter{\romannumeral1} and case \uppercase\expandafter{\romannumeral2}, respectively.
The start points of these two cases both locate at the $\mathcal{PT}$-symmetry broken phase, and the encircling directions are all clockwise.
Considering the Hamiltonian of the NV center, we multiply an overall coefficient of $10^{-3}~\mathrm{GHz}$ to the $H_s$  in equation (1) of the main text.
The inner products of the encircling state, $\langle\psi(t)|\psi(t)\rangle$, are given in Supplementary Fig.~\ref{Necessity_of_the_Indirect_Measurement}A, where we can see that the inner products cover several orders of magnitude for both cases.
The initial operators of $M(t)$ were both set as $M(0)=1.5\cdot I$ when dilating $H_s(t)$ into $H_{s,a}(t)$.
The real value functions $b(t)$ were both chosen as the form given in Supplementary Fig.~\ref{Necessity_of_the_Indirect_Measurement}B to keep operator $M(t)-I$ positive during the dilation process.
The probabilities of obtaining the ancilla qubit state $|-\rangle$, $P_-(t)$, are plotted in Supplementary Fig.~\ref{Necessity_of_the_Indirect_Measurement}C for both cases.
Supplementary Fig.~\ref{Necessity_of_the_Indirect_Measurement}C shows that the $P_-(t)$ in case \uppercase\expandafter{\romannumeral1} is big enough for measurement all the time.
However, the $P_-(t)$ in case \uppercase\expandafter{\romannumeral2} is only the magnitude of $10^{-4}$ for some time and many repeat times of the experiment will be needed to measure it.

\newpage
\begin{figure}[htbp]
\centering
\includegraphics[width=15cm]{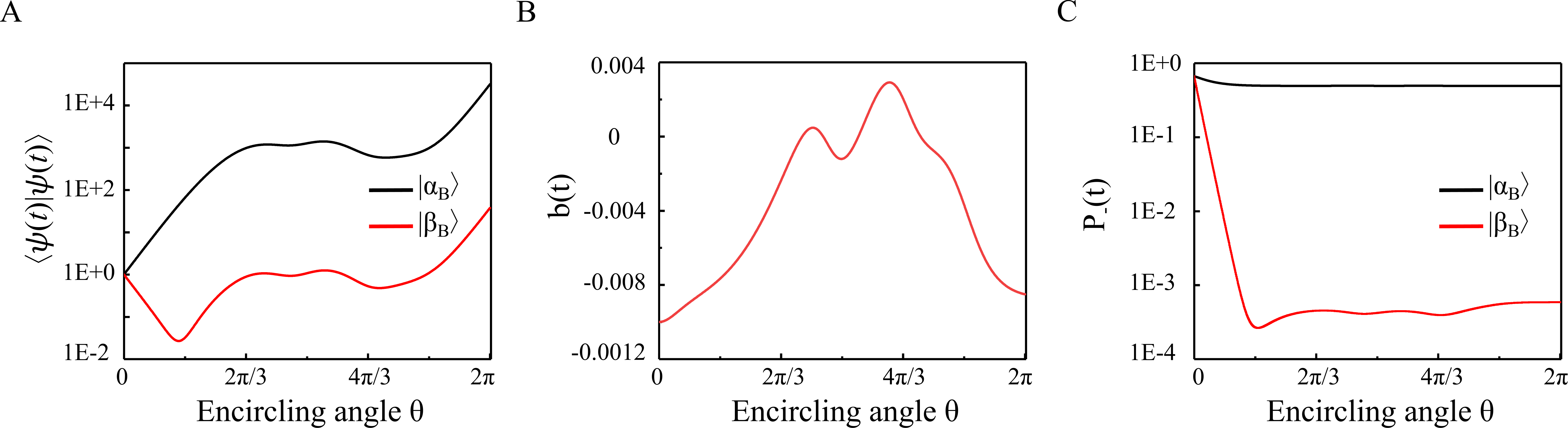}
\caption{\textbf{Difficulty for direct measurements.}
 (\textbf{A}) Inner product of the encircling state when the initial state is $|\alpha_B\rangle$ (black line, case \uppercase\expandafter{\romannumeral1}) and $|\beta_B\rangle$ (red line, case \uppercase\expandafter{\romannumeral2}), respectively. (\textbf{B}) real value function $b(t)$ versus encircling angle $\theta$. (\textbf{C}) Probability of obtaining ancilla qubit state $|-\rangle$, $P_-(t)$, when the initial state is $|\alpha_B\rangle$ (black line) and $|\beta_B\rangle$ (red line), respectively. Logarithmic coordinates are chosen in the vertical axis of (A) and (C). }
\label{Necessity_of_the_Indirect_Measurement}
\end{figure}

To reduce the difficulty of the experiment, we therefore introduce the measurement method in the main text.
The dilated state of the two-qubit system in equation \ref{Dilated State} can also be written as
\begin{equation}
\label{Dilated_Encircling_State}
|\Psi(t)\rangle=\frac{[I-i\eta(t)]|\psi(t)\rangle\otimes|0\rangle-i[I+i\eta(t)]|\psi(t)\rangle\otimes|1\rangle}
{\sqrt{\langle\psi(t)|\eta^\dagger(t)\eta(t)+I|\psi(t)\rangle}}.
\end{equation}
The state of the system qubit when the ancilla qubit state is $|0\rangle$ is defined as
\begin{equation}
\label{State_Chi}
|\chi(t)\rangle=\frac{[I-i\eta(t)]|\psi(t)\rangle}{\sqrt{\langle\psi(t)|\eta^\dagger(t)\eta(t)+I|\psi(t)\rangle}}.
\end{equation}
We use quantum state tomography to measure state $|\chi(t)\rangle$.
Then state $|\psi(t)$ can be obtained by multiplying $[I-i\eta(t)]^{-1}$ to state $|\chi(t)\rangle$, renormalizing the state and finally using the maximum likelihood estimation \cite{PRA_Fiurasek_SM}.

%#####%  Experimental obtain of the encircling state---PL Rate
The four levels give rise to different photoluminescence(PL) rates \cite{PRB_Steiner_SM}, labeled by $L_{|0\rangle_e|1\rangle_n}$, $L_{|0\rangle_e|0\rangle_n}$, $L_{|-1\rangle_e|1\rangle_n}$, and $L_{|-1\rangle_e|0\rangle_n}$, respectively.
The PL rate differences can be used to readout the state of the two-qubit system.
However, states with different population distributions can display the same PL rate, thus a set of pulse sequences given in Supplementary Fig.~\ref{Quantum_State_Tomography_Pulse_Sequence} were utilized to measure the dilated state.

\newpage
\begin{figure}[htbp]
\centering
\includegraphics[width=15cm]{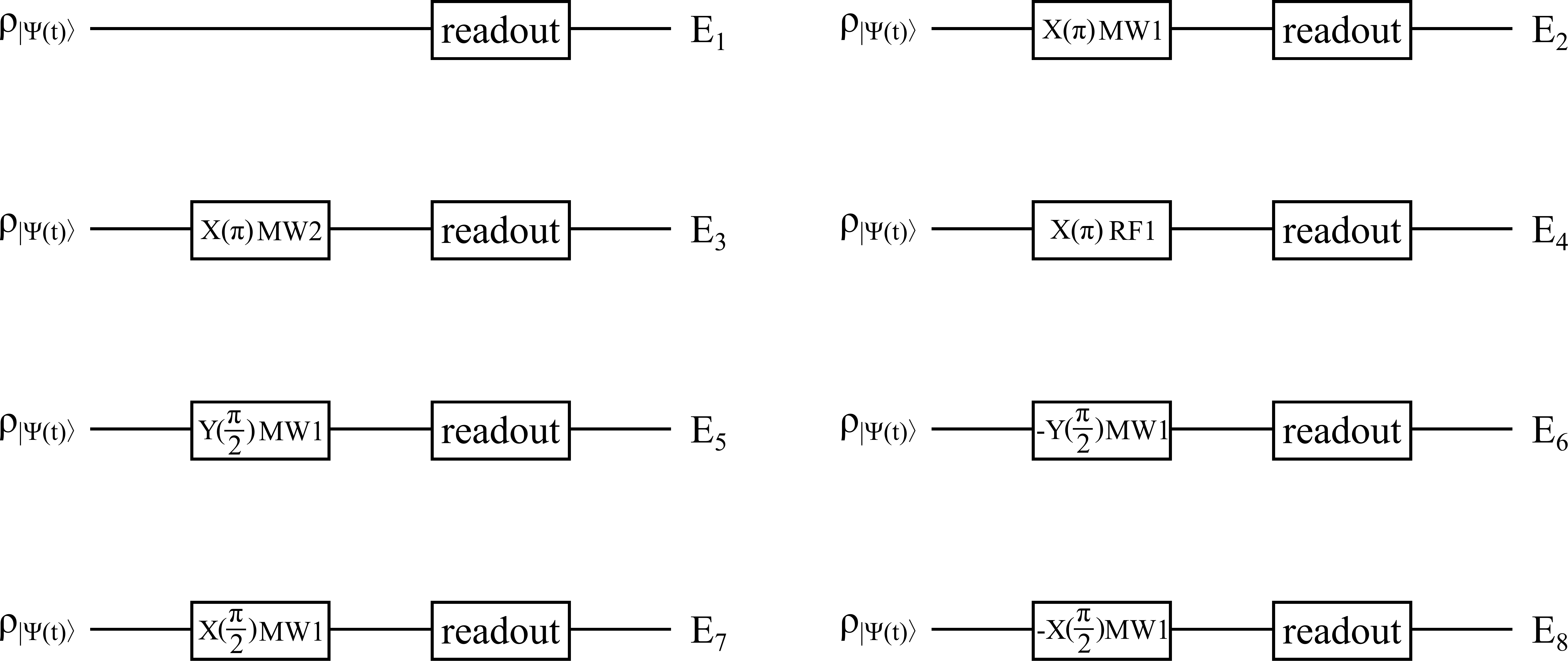}
\caption{
\textbf{Pulse Sequences for Quantum State Tomography.}
$\rho_{|\Psi(t)\rangle}$ denotes the dilated state.
$\mathrm{X(\pi)~MW1}$ in the second pulse sequence denotes the selective MW pulse between level $|0\rangle_e|1\rangle_n$ and level $|-1\rangle_e|1\rangle_n$ (see Fig.~2 in the main text) along the X-axis, and the rotate angle is $\pi$. For other pulse sequences, the definitions are similar.
$E_i$ represents the obtained PL rate of the state after the effect of the $i^{th}$ pulse sequence.
  }
\label{Quantum_State_Tomography_Pulse_Sequence}
\end{figure}

%#####%  Experimental obtain of the encircling state---Quantum State Tomography
The density matrix of the dilated state $|\Psi(t)\rangle$ can be denoted by
\begin{equation}
\label{Density_Matrix_of_Dilated_Encirlcing_State}
\rho_{|\Psi(t)\rangle}=\left(
       \begin{array}{cccc}
         \rho_{11} & \rho_{12} & \rho_{13} & \rho_{14} \\
         \rho_{21} & \rho_{22} & \rho_{23} & \rho_{24} \\
         \rho_{31} & \rho_{32} & \rho_{33} & \rho_{34} \\
         \rho_{41} & \rho_{42} & \rho_{43} & \rho_{44} \\
      \end{array}
     \right),
\end{equation}
which means the density matrix of state $|\chi(t)\rangle$ is
\begin{equation}
\label{Density_Matrix_of_chi}
\rho_{|\chi(t)\rangle}=\left(
                      \begin{array}{cc}
                        \rho_{11} & \rho_{13} \\
                        \rho_{31} & \rho_{33} \\
                      \end{array}
                    \right).
\end{equation}
The first pulse sequence in Supplementary Fig.~\ref{Quantum_State_Tomography_Pulse_Sequence} represents direct readout of state $|\Psi(t)\rangle$, and the obtained PL rate was
\begin{equation}
\label{PL_Rate_Example}
E_1=\rho_{11}L_{|0\rangle_e|1\rangle_n}+\rho_{22}L_{|0\rangle_e|0\rangle_n}
+\rho_{33}L_{|-1\rangle_e|1\rangle_n}+\rho_{44}L_{|-1\rangle_e|0\rangle_n}.
\end{equation}
Similarly, we can write down the PL rates of the states after the effect of other pulses, this yields
\begin{equation}
\label{PL_Equations}
\begin{bmatrix}
\rho_{11} & \rho_{22} & \rho_{33} & \rho_{44} \\
\rho_{33} & \rho_{22} & \rho_{11} & \rho_{44} \\
\rho_{11} & \rho_{44} & \rho_{33} & \rho_{22} \\
\rho_{22} & \rho_{11} & \rho_{33} & \rho_{44} \\
\frac{\rho_{11}+\rho_{33}}{2} -\frac{(\rho_{13}+\rho_{31})}{2}     & \rho_{22} & \frac{\rho_{11}+\rho_{33}}{2}+\frac{(\rho_{13}+\rho_{31})}{2}     & \rho_{44}     \\
\frac{\rho_{11}+\rho_{33}}{2}+\frac{(\rho_{13}+\rho_{31})}{2}     & \rho_{22} & \frac{\rho_{11}+\rho_{33}}{2} -\frac{(\rho_{13}+\rho_{31})}{2}     & \rho_{44}     \\
\frac{\rho_{11}+\rho_{33}}{2} -\frac{(\rho_{13}-\rho_{31})}{2}     & \rho_{22} & \frac{\rho_{11}+\rho_{33}}{2}+\frac{(\rho_{13}-\rho_{31})}{2}     & \rho_{44}     \\
\frac{\rho_{11}+\rho_{33}}{2}+\frac{(\rho_{13}-\rho_{31})}{2}     & \rho_{22} & \frac{\rho_{11}+\rho_{33}}{2} -\frac{(\rho_{13}-\rho_{31})}{2}     & \rho_{44}     \\
\end{bmatrix}
\begin{bmatrix}
L_{|0\rangle_e|1\rangle_n} \\ L_{|0\rangle_e|0\rangle_n} \\ L_{|-1\rangle_e|1\rangle_n} \\ L_{|-1\rangle_e|0\rangle_n} \\
\end{bmatrix}
=
\begin{bmatrix}
E_1 \\ E_2 \\ E_3 \\ E_4 \\  E_5 \\  E_6 \\  E_7 \\  E_8
\end{bmatrix}.
\end{equation}
By solving equation \ref{PL_Equations}, we can obtain that the elements of the density matrix $\rho_{|\chi(t)\rangle}$ are
\begin{equation}
\label{Solution_of_PL_Equations}
\left\{
\begin{aligned}
\rho_{11}&=\frac{1}{4}+\frac{E_1-E_4}{2(L_{|0\rangle_e|1\rangle_n}-L_{|0\rangle_e|0\rangle_n})}
+\frac{E_1-E_2}{4(L_{|0\rangle_e|1\rangle_n}-L_{|-1\rangle_e|1\rangle_n})}
+\frac{E_1-E_3}{4(L_{|0\rangle_e|0\rangle_n}-L_{|-1\rangle_e|0\rangle_n})}, \\
\rho_{33}&=\frac{1}{4}+\frac{E_1-E_4}{2(L_{|0\rangle_e|1\rangle_n}-L_{|0\rangle_e|0\rangle_n})}
-\frac{3(E_1-E_2)}{4(L_{|0\rangle_e|1\rangle_n}-L_{|-1\rangle_e|1\rangle_n})}
+\frac{E_1-E_3}{4(L_{|0\rangle_e|0\rangle_n}-L_{|-1\rangle_e|0\rangle_n})},  \\
\rho_{13}&=\frac{E_6-E_5+i(E_8-E_7)}{2(L_{|0\rangle_e|1\rangle_n}-L_{|-1\rangle_e|1\rangle_n})}, \\
\rho_{31}&=\frac{E_6-E_5-i(E_8-E_7)}{2(L_{|0\rangle_e|1\rangle_n}-L_{|-1\rangle_e|1\rangle_n})},
\end{aligned}
\right.
\end{equation}
where $L_{|0\rangle_e|1\rangle_n}-L_{|-1\rangle_e|1\rangle_n}$, $L_{|0\rangle_e|1\rangle_n}-L_{|0\rangle_e|0\rangle_n}$, and $L_{|0\rangle_e|0\rangle_n}-L_{|-1\rangle_e|0\rangle_n}$ denote the PL rate differences between the corresponding energy levels.
%$|0\rangle_e|1\rangle_n$ and level $|-1\rangle_e|1\rangle_n$, level $|0\rangle_e|1\rangle_n$ and level $|0\rangle_e|0\rangle_n$, level $|0\rangle_e|0\rangle_n$ and level $|-1\rangle_e|0\rangle_n$, respectively.
These PL rate differences were measured by Rabi oscillation experiments.

%#####%  Experimental obtain of the encircling state---Transform From State Chi to State Psi
With equation \ref{State_Chi}, the density matrix of state $|\psi(t)\rangle$ can be calculated by
\begin{equation}
\label{State_Chi_to_State_Psi}
\rho_{|\psi(t)\rangle}=c(t) [I-i\eta(t)]^{-1}\rho_{|\chi(t)\rangle}\{[I-i\eta(t)]^{-1}\}^\dag,
\end{equation}
where $c(t)$ is a normalization coefficient, and the operator $\eta(t)$ can be obtained from equation \ref{Relation of operator  M and eta}, that is $\eta(t)=\sqrt{M(t)-I}$.

%#####%  Experimental obtain of the encircling state---Maximum Likely Estimation
The obtained density matrix $\rho_{|\psi(t)\rangle}$ may be non-physical.
To solve this problem, a maximum likelihood estimation method is utilized to find the most possible physical state as the final $|\psi(t)\rangle$.

\begin{figure}[htbp]
\centering
\includegraphics[width=15cm]{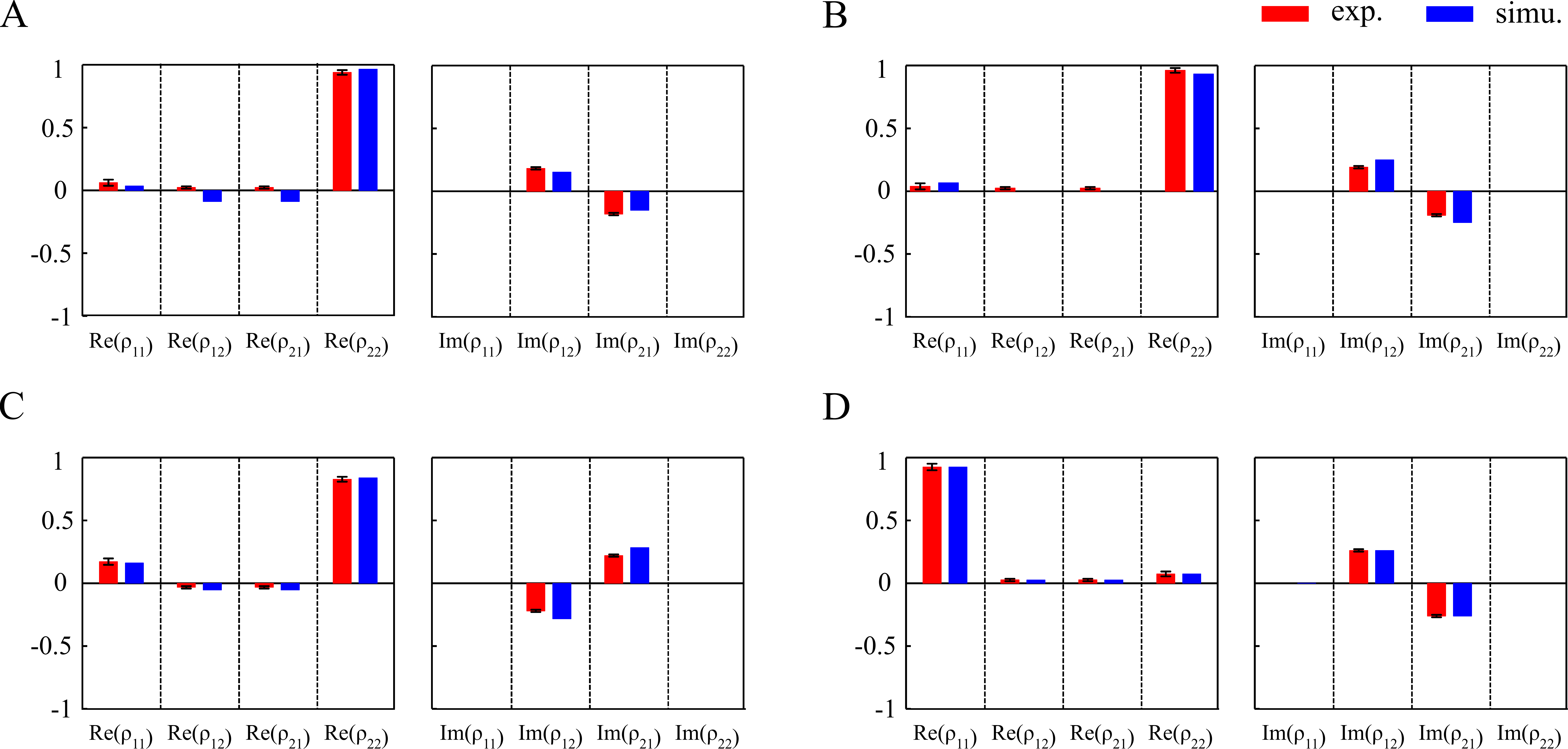}
\caption{\textbf{Quantum State Tomography and Transformation.}
 (\textbf{A-B}) Result of the encircling initial state.
Measured density matrix $\rho_{|\chi(0)\rangle}$ (A) and transformed density matrix $\rho_{|\psi(0)\rangle}$ (B).
(\textbf{C-D}) Result of the encircling final state.
  Measured density matrix $\rho_{|\chi(2\pi)\rangle}$ (C) and transformed state $\rho_{|\psi(2\pi)\rangle}$ (D).
 The left panels are the real parts of these density matrixes while the right panels show the image parts.
 The red bars are the experimental results and the blue bars are the simulation results.
  }
\label{Quantum_State_Tomography_Density_Matrix_Histogram}
\end{figure}

%#####%  Experimental obtain of the encircling state---Experiment result
Supplementary Fig.~\ref{Quantum_State_Tomography_Density_Matrix_Histogram} shows the results of the measured density matrix $\rho_{|\chi(\theta)\rangle}$ and the transformed density matrix $\rho_{|\psi(\theta)\rangle}$ by taking the case \uppercase\expandafter{\romannumeral2} as an example (Here we have replaced $\theta$ for t and the relation is $\theta=2\pi t/T$).
The results of the encircling initial state are displayed in Supplementary Fig.~\ref{Quantum_State_Tomography_Density_Matrix_Histogram}A and B.
%Fig~.\ref{Quantum_State_Tomography_Density_Matrix_Histogram}A shows the results of the measured density matrix $\rho_{|\chi(0)\rangle}$ while Fig~.\ref{Quantum_State_Tomography_Density_Matrix_Histogram}B shows the results of the transformed density matrix $\rho_{|\psi(0)\rangle}$.
The red bars are the experimental results and the blue bars are the simulation results which included the dephasing noise of the electron spin.
% and the crosstalk of these MW pulses.
The fidelities between the experimental obtained density matrixes and the simulated ones in Supplementary Fig.~\ref{Quantum_State_Tomography_Density_Matrix_Histogram}A and B are $F_{\rho_{|\chi(0)\rangle}}=0.98(3)$ and $F_{\rho_{|\psi(0)\rangle}}=1.00(4)$, respectively.
As for the encircling final state, the measured density matrix $\rho_{|\chi(2\pi)\rangle}$ is depicted in Supplementary Fig.~\ref{Quantum_State_Tomography_Density_Matrix_Histogram}C and the transformed density matrix $\rho_{|\psi(2\pi)\rangle}$ is given in Supplementary Fig.~\ref{Quantum_State_Tomography_Density_Matrix_Histogram}D.
The fidelities between the experimental and simulated density matrixes in Supplementary Fig.~\ref{Quantum_State_Tomography_Density_Matrix_Histogram}C and D are $F_{\rho_{|\chi(0)\rangle}}=0.99(1)$ and $F_{\rho_{|\psi(0)\rangle}}=1.00(4)$, respectively.
For other encircling angles and each encircling cases, the fidelities between the experimental and simulated results are summarized in the Supplementary Table \ref{Fidelity_Results}.
The average fidelity between the experimental results and the simulated results for the measured density matrix $\rho_{|\chi(\theta)\rangle}$ and the transformed density matrix $\rho_{|\psi(\theta)\rangle}$ are both close to 1.
These high fidelities show that the measured states agree well with the simulation predications.

\begin{table}
\centering
\caption
{\textbf{Summarization of the fidelities.}
For each encircling cases and different encircling angles, the fidelity between the measured density matrix $\rho_{|\chi(\theta)\rangle}$ and the simulation result is denoted by $F_{\rho_{|\chi(t)\rangle}}$, while $F_{\rho_{|\psi(t)\rangle}}$ represents the fidelity between the transformed density matrix $\rho_{|\psi(\theta)\rangle}$ and the corresponding simulation result.
The average fidelity is $\bar{F}_{\rho_{|\chi(\theta)\rangle}}=0.98(2)$ and  $\bar{F}_{\rho_{|\psi(\theta)\rangle}}=1.00(4)$, respectively.
}
\textrm{\\}
\renewcommand{\multirowsetup}{\centering}
\begin{tabular}{|c|c|c|c|c|c|c|c|c|}            % {lccc} 表示各列元素对齐方式，left-l,right-r,center-c
\hline
\multirow{3}{*}                                                              %  这一行分三小行
%{\diagbox[width=3.5cm,trim=l]{$\theta$}{Fidelity}{cases}}%   第一小行第一列占三小行
{\diagbox{}{}}
& \multicolumn{4}{c|}{start point A}                          %   第一小行第二列包含四小列     中间对齐最右加线
& \multicolumn{4}{c|}{start point B}   \\                     %   第一小行结束加换行符  \\
\cline{2-9}                                                                       %   第一小行分隔线
& \multicolumn{2}{c|}{$\circlearrowright$}              %   第二小行第二列包含两小列
& \multicolumn{2}{c|}{$\circlearrowleft$}
& \multicolumn{2}{c|}{$\circlearrowright$}
& \multicolumn{2}{c|}{$\circlearrowleft$}  \\           %  第二小行结束加换行符  \\
\cline{2-9}                                                                      %  第二小行分隔线
&$|\alpha_A\rangle$&$|\beta_A\rangle$&$|\alpha_A\rangle$&$|\beta_A\rangle$
&$|\alpha_B\rangle$&$|\beta_B\rangle$&$|\alpha_B\rangle$&$|\beta_B\rangle$    \\
%&\multicolumn{3}{c|}{}&50&100&150&200&50    \\
\hline\hline
$F_{\rho_{|\chi(0)\rangle}}$         &0.98(4) &1.00(5) &1.00(2) &0.99(3) &1.00(4) &0.97(3) &1.00(4) &0.98(3)   \\
\hline
$F_{\rho_{|\psi(0)\rangle}}$          &1.00(5) &1.00(5) &1.00(5) &1.00(5) &0.99(4) &1.00(4) &0.99(4) &1.00(4)    \\
\hline\hline
$F_{\rho_{|\chi(\pi/3)\rangle}}$    &0.96(3) &0.98(4) &0.96(3) &0.97(3) &0.97(2) &0.99(2) &0.99(1) &0.99(3)    \\
\hline
$F_{\rho_{|\psi(\pi/3)\rangle}}$    &1.00(3) &1.00(4) &1.00(4) &1.00(3) &1.00(3) &1.00(4) &1.00(4) &1.00(4)   \\
\hline\hline
$F_{\rho_{|\chi(2\pi/3)\rangle}}$  &0.99(3) &0.99(3) &0.98(3) &0.96(2) &0.99(1) &0.98(1) &1.00(1) &0.99(3)   \\
\hline
$F_{\rho_{|\psi(2\pi/3)\rangle}}$  &1.00(4) &0.99(4) &0.98(4) &1.00(4) &1.00(4) &1.00(4) &1.00(4) &1.00(4)    \\
\hline\hline
$F_{\rho_{|\chi(\pi)\rangle}}$       &0.96(1) &0.99(3) &0.97(4) &0.98(1) &0.97(1) &1.00(1) &0.99(1) &0.96(3)    \\
\hline
$F_{\rho_{|\psi(\pi)\rangle}}$       &1.00(3) &1.00(4) &1.00(5) &1.00(4) &1.00(3) &1.00(4) &1.00(4) &1.00(3)   \\
\hline\hline
$F_{\rho_{|\chi(4\pi/3)\rangle}}$  &1.00(1) &0.97(3) &0.99(1) &0.97(1) &0.99(1) &0.99(5) &0.97(2) &0.98(3)    \\
\hline
$F_{\rho_{|\psi(4\pi/3)\rangle}}$  &1.00(4) &1.00(4) &1.00(4) &1.00(3) &1.00(5) &1.00(5) &1.00(4) &1.00(4)  \\
\hline\hline
$F_{\rho_{|\chi(5\pi/3)\rangle}}$  &0.99(2) &0.99(1) &0.99(1) &0.98(1) &0.98(2) &0.99(1) &0.98(1) &0.98(2)   \\
\hline
$F_{\rho_{|\psi(5\pi/3)\rangle}}$  &1.00(4) &1.00(4) &1.00(4) &1.00(4) &1.00(4) &1.00(3) &1.00(3) &1.00(3)   \\
\hline\hline
$F_{\rho_{|\chi(2\pi)\rangle}}$     &0.99(1) &1.00(1) &0.99(3) &0.99(1) &0.98(1) &0.99(1) &0.99(1) &0.99(1)   \\
\hline
$F_{\rho_{|\psi(2\pi)\rangle}}$     &1.00(4) &1.00(5) &1.00(5) &1.00(3) &1.00(3) &1.00(4) &1.00(3) &1.00(4)    \\
\hline
\end{tabular}
\label{Fidelity_Results}
\end{table}

\section{Note \uppercase\expandafter{\romannumeral4}: Construction of $H_{s,a}(t)$ in the NV Center}

%#####%  Construct the Dilated Hamiltonian---Description
\quad ~This part demonstrates the construction of the dilated Hamiltonian $H_{s,a}(t)$ in the NV center.

The dilated Hamiltonian we need to construct is
\begin{equation}
\label{Decomposited_Dilated_Hamiltonian}
H_{s,a}(t)=\sum_{i=0}^3[ A_i(t) \sigma_i\otimes I+B_i(t)\sigma_i\otimes \sigma_z].
\end{equation}
The static Hamiltonian of the NV center has the form
\begin{equation}
\label{NV_Hamiltonian_Allspace}
H_{\mathrm{NV}}=2\pi(DS_z^2 + \omega_eS_z + QI_z^2 + \omega_nI_z + AS_zI_z).
\end{equation}
The subspace spanned by $|0\rangle_e|1\rangle_n$, $|0\rangle_e|0\rangle_n$, $ |-1\rangle_e|1\rangle_n$ and $|-1\rangle_e|0\rangle_n$ is utilized to form a two-qubit system to perform the experiment.
The static Hamiltonian of the NV center in this subspace can be simplified as
\begin{equation}
\label{NV_Hamiltonian_Subspace}
H_0=\pi [-(D-\omega_e-\frac{A}{2})\sigma_z\otimes I+(Q+\omega_n-\frac{A}{2})I\otimes\sigma_z+\frac{A}{2}\sigma_z\otimes\sigma_z].
%+\pi(D-\omega_e-Q-\omega_n+\frac{A}{2})I\otimes I.
\end{equation}
To construct $H_{s,a}(t)$ in this subspace, we apply two slightly detuned MW pulses to selectively drive the electron spin transitions, as depicted in Fig.~2b in the main text.
The Hamiltonian of the pulses can be written as
\begin{equation}
\label{MW_Hamiltonian}
\begin{aligned}
H_C(t)&=\Omega_1(t)\cos[\int_0^t\omega_1(\tau)d\tau+\phi_1(t)]\sigma_x\otimes|1\rangle_n~_n\langle1|\\ &+\Omega_2(t)\cos[\int_0^t\omega_2(\tau)d\tau+\phi_2(t)]\sigma_x\otimes|0\rangle_n~_n\langle0|.
\end{aligned}
\end{equation}
where $\Omega_1(t)$, $\omega_1(t)$ and $\phi_1(t)$ ($\Omega_2(t)$, $\omega_2(t)$ and $\phi_2(t)$) are the amplitude, angular frequency and  phase of the MW1 (MW2) pulse.
Thus the total Hamiltonian of the NV center when applying MW pulses is
\begin{equation}
\label{NV_Hamiltonian_Total}
\begin{aligned}
H_{\mathrm{tot}}(t)=& \pi [-(D-\omega_e-\frac{A}{2})\sigma_z\otimes I+(Q+\omega_n-\frac{A}{2})I\otimes\sigma_z+\frac{A}{2}\sigma_z\otimes\sigma_z]  \\
&+\Omega_1(t)\cos[\int_0^t\omega_1(\tau)d\tau+\phi_1(t)]\sigma_x\otimes|1\rangle_n~_n\langle1|\\ &+\Omega_2(t)\cos[\int_0^t\omega_2(\tau)d\tau+\phi_2(t)]\sigma_x\otimes|0\rangle_n~_n\langle0|.
\end{aligned}
\end{equation}
To construct $H_{s,a}(t)$ in the NV center, we need to select an appropriate interaction picture and  delicately set the parameters $\Omega_1(t)$, $\Omega_2(t)$, $\omega_1(t)$, $\omega_2(t)$, $\phi_1(t)$ and $\phi_2(t)$.

%#####%  Construct the Dilated Hamiltonian---Solution
Comparing the diagonal components of $H_{s,a}(t)$ and $H_{tot}(t)$, we can choose the interaction picture as
\begin{equation}
\label{Interaction Picture}
U_{rot}(t)=e^{i\int_0^t[H_0-A_0(\tau)I\otimes I-A_3(\tau)\sigma_z\otimes I-B_0(\tau)I\otimes\sigma_z-B_3(\tau)\sigma_z\otimes\sigma_z]d\tau}.
\end{equation}
The total Hamiltonian of the NV center in this interaction picture then transforms to
\begin{equation}
\label{Hamiltonian_in_Interaction_Picture}
\begin{aligned}
H_{rot}(t)=& U_{rot}(t)H_{tot}(t)U_{rot}^\dag(t)-iU_{rot}(t)\frac{dU_{rot}^\dag(t)}{dt}                \\ =&A_0(\tau)I\otimes I+A_3(\tau)\sigma_z\otimes                                I+B_0(\tau)I\otimes\sigma_z+B_3(\tau)\sigma_z\otimes\sigma_z      \\
&+\Omega_1(t)\cos[\int_0^t\omega_1(\tau)d\tau+\phi_1(t)]
\{\cos[\omega_{MW1}t+\int_0^t[2A_3(\tau)+2B_3(\tau)]d\tau]\sigma_x \\
&+\sin[\omega_{MW1}t+\int_0^t[2A_3(\tau)+2B_3(\tau)]d\tau]\sigma_y\}\otimes|1\rangle_n~_n\langle1|\\
&+\Omega_2(t)\cos[\int_0^t\omega_2(\tau)d\tau+\phi_2(t)]
\{\cos[\omega_{MW2}t+\int_0^t[2A_3(\tau)-2B_3(\tau)]d\tau]\sigma_x\\
&+\sin[\omega_{MW2}t+\int_0^t[2A_3(\tau)-2B_3(\tau)]d\tau]\sigma_y\}\otimes|0\rangle_n~_n\langle0|,
\end{aligned}
\end{equation}
with $\omega_{MW1}=2\pi(D-\omega_e-A)$ ($\omega_{MW2}=2\pi(D-\omega_e)$) being the transition frequency between $|0\rangle_e|1\rangle_n$ and $|-1\rangle_e|1\rangle_n$ ($|0\rangle_e|0\rangle_n$ and $|-1\rangle_e|0\rangle_n$).
By choosing
\begin{equation}
\label{Frequency_of MW_Pulses}
\left\{
\begin{aligned}
&\omega_1(t)=\omega_{MW1}+2A_3(t)+2B_3(t),   \\
&\omega_2(t)=\omega_{MW2}+2A_3(t)-2B_3(t),   \\
\end{aligned}
\right.
\end{equation}
and in the condition of rotating wave approximation, we can simplify $H_{rot}(t)$ as
\begin{equation}
\label{Hamiltonian_in_Interaction_Picture_Simplification}
\begin{aligned}
H_{rot}(t)
%=&A_0(\tau)I\otimes I+A_3(\tau)\sigma_z\otimes                                I+B_0(\tau)I\otimes\sigma_z+B_3(\tau)\sigma_z\otimes\sigma_z      \\
%&+\pi\Omega_1(t)[\cos\phi_1(t)\sigma_x-\sin\phi_1(t)\sigma_y]\otimes|1\rangle_n~_n\langle1|    \\
%&+\pi\Omega_2(t)[\cos\phi_2(t)\sigma_x-\sin\phi_2(t)\sigma_y]\otimes|0\rangle_n~_n\langle0|    \\
%=&A_0(\tau)I\otimes I+A_3(\tau)\sigma_z\otimes                                I+B_0(\tau)I\otimes\sigma_z+B_3(\tau)\sigma_z\otimes\sigma_z      \\
%&+\sigma_x\otimes[\pi\Omega_1(t)\cos\phi_1(t)|1\rangle_n~_n\langle1|
%+\pi\Omega_2(t)\cos\phi_2(t)|0\rangle_n~_n\langle0|] \\
%&-\sigma_y\otimes[\pi\Omega_1(t)\sin\phi_1(t)|1\rangle_n~_n\langle1|
%+\pi\Omega_2(t)\sin\phi_2(t)|0\rangle_n~_n\langle0|],  \\
=&A_0(\tau)I\otimes I+A_3(\tau)\sigma_z\otimes                                I+B_0(\tau)I\otimes\sigma_z+B_3(\tau)\sigma_z\otimes\sigma_z      \\
&+\frac{\Omega_1(t)\cos\phi_1(t)+\Omega_2(t)\cos\phi_2(t)}{4}\sigma_x\otimes I
+\frac{\Omega_1(t)\cos\phi_1(t)-\Omega_2(t)\cos\phi_2(t)}{4}\sigma_x\otimes \sigma_z \\
&+\frac{\Omega_1(t)\sin\phi_1(t)+\Omega_2(t)\sin\phi_2(t)}{4}\sigma_y\otimes I
+\frac{\Omega_1(t)\sin\phi_1(t)-\Omega_2(t)\sin\phi_2(t)}{4}\sigma_y\otimes \sigma_z  .
\end{aligned}
\end{equation}
Comparing equation \ref{Hamiltonian_in_Interaction_Picture_Simplification} with equation \ref{Decomposited_Dilated_Hamiltonian}, if we choose
\begin{equation}
\label{Rabi_Frequencies_and_Phases_of_MW_Pulses}
\left\{
\begin{aligned}
&\Omega_1(t)=2\sqrt{[A_1(t)+B_1(t)]^2+[A_2(t)+B_2(t)]^2},  \\
&\Omega_2(t)=2\sqrt{[A_1(t)-B_1(t)]^2+[A_2(t)-B_2(t)]^2},    \\
&\phi_1(t)=-\mathrm{atan2}[A_2(t)+B_2(t),A_1(t)+B_1(t)],       \\
&\phi_2(t)=-\mathrm{atan2}[A_2(t)-B_2(t),A_1(t)-B_1(t)],
\end{aligned}
\right.
\end{equation}
then the dilated Hamiltonian $H_{s,a}(t)$ can be realized.

\end{document}